
\def\midfig#1x#2:#3#4{\midinsert
$$\vbox to #2{\hbox to #1{\special{ps:#3}\hfill}\vfill}$$\par
#4\par\endinsert}
\def\topfig#1x#2:#3#4{\topinsert
$$\vbox to #2{\hbox to #1{\special{ps:#3}\hfill}\vfill}$$\par
#4\par\endinsert}
\def\infig#1x#2:#3{
$$\vbox to #2{\hbox to #1{\special{ps:#3}\hfill}\vfill}$$}
\def\textfig#1x#2:#3{
$\vbox to #2{\hbox to #1{\special{ps:#3}\hfill}\vfill}$}
\input definit.tex
\magnification=1200
\baselineskip=22.0 truept
\centerline{{\bf Time Invariant Scaling in Discrete Fragmentation Processes}}

\bigskip
\centerline{B.G.Giraud and R.Peschanski}
\centerline{\it{Service Physique Th\'eorique, DSM-CE Saclay, F91191
Gif/Yvette, France}}

\bigskip
Abstract: Linear rate equations are used to describe the cascading decay of
an initial heavy cluster into fragments. We consider moments of arbitrary
orders of the mass multiplicity spectrum and derive scaling properties
pertaining to their time evolution. We suggest that the mass weighted
multiplicity is a suitable observable for the discovery of scaling.
Numerical tests validate such properties, even
for moderate values of the initial mass
(nuclei, percolation clusters, etc.). Finite size effects can be simply
parametrized.

\bigskip
In this work we consider binary fragmentation processes where any
fragment with mass number $k$ breaks into fragments with mass numbers
$j$ and $k-j,$ $j =1,2...k-1,$ with a probability $w_{j k}$
per unit of time.
It is assumed that $w_{j k}$ is time independent. By definition, $w_{jk}=0$
if $j \ge k$ and
$w_{j k}$
is symmetric if $j$ is replaced by $k-j,$ naturally. For technical reasons, we
will use for $w_{jk}$ the double of the actual transition rate whenever the
special case $w_{j,2j}$ occurs. Let
$N_j(t)$ be the multiplicity of fragment $j$ at time $t$ in a process initiated
from the decay of a cluster $A,$ namely $N_j(0)=\delta_{jA}.$
The model under study is described by the following set of linear,
first order differential equations,
$$
{dN_j \over dt}= - c_j N_j + \sum_{k=j+1}^A w_{jk} N_k,\ j=1,...A, \eqno (1)
$$
with
$$
c_j=\sum_{\ell=1}^{j-1} {w_{\ell j} \over 2}. \eqno (2)
$$
With components $N_j,\ j=1,...A,$ for a column vector
$|{\cal N}>,$ the system, Eqs.(1),
boils down to
$d|{\cal N}> / dt = {\cal W |N>}$ with a triangular matrix $\cal W.$
The general solution of Eqs.(1) is obviously a sum of exponentials whose rates
of decay in time are the trivial
eigenvalues of the triangular $\cal W,$
namely the diagonal matrix elements $-c_j.$

\medskip
This matrix $\cal W$ has a remarquable property, namely a fixed left
(row-like) eigenstate ${\cal M}_1$, whose components
${\cal M}_{1j}=j,\ j=1,...A,$ do not depend on the $w_{\ell k}$'s.
This comes from the symmetries of $\cal W$ demanded by the conservation of
the total mass $M_1=\sum_{j=1}^A j N_j$, with
$dM_1 / dt=<{\cal M}_1|{\cal W}|{\cal N}>.$
The corresponding eigenvalue
is, naturally, $-c_1=0.$ It is thus convenient to define the ``mass
weighted multiplicity'' (MWM) vector $U$
with components $U_i=i N_i,$ whose evolution is governed by a matrix
${\cal V}$, with matrix elements ${\cal V}_{jk}=j{\cal W}_{jk}/k,$ hence
$$
dU/dt={\cal V}U. \eqno (3)
$$
\noindent
The surface under the histogram defined by $U$ is therefore time invariant,
$\sum_{i=1}^A U_i/A = 1.$ This invariant normalization gives a hint that
an analysis of $U$ might be the best way to reveal scaling and even
universality properties of the fragmentation process.

\medskip
For that analysis, we first consider a moment of arbitrary order $q$
$$
M_q=\sum_{j=1}^A j^{q-1} U_j=\sum_{j=1}^A j^q N_j, \eqno (4)
$$
where the exponent $q$ is any real number.
The time derivative of such a moment is
$$
{dM_q \over dt}= \sum^A_{j=1} j^q \left( \sum^
A_{k=j+1}w_{jk}N_k-c_jN_j \right)= \sum^ A_{k=1} k^{q-1} U_k d(q,k), \eqno (5)
$$
with
$$
d(q,k)= \sum^{k-1}_{j=1} w_{jk} \left[ \left({j \over k}
\right)^q - {1 \over 2} \right], \eqno (6)
$$
where we have used an interchange of indices $j$ and $k$ in the
double summation.

\medskip
The structure of the right-hand-side of Eq.(5) gives a special interest
to those cases where $d(q,k)$ shows scaling properties with respect to $k.$
For instance$^{1)}$ if
a condition $ d(q,k) \simeq d(q) $ is met, this approximate
independence with respect to $k$ induces an eigenvalue problem where the
corresponding (eigen)moments $M_q$ are decoupled and
decay exponentially with respect to time. A sufficient condition for this is
a behavior
$w_{jk} \simeq k^{-1}\varphi(j/k),$
whence, when $k$ is large enough,
$$
d(q,k) = {1\over k} \sum^{k-1}_{j=1} \varphi(j/k) \left[ \left({j \over k}
\right)^q - {1 \over 2} \right] \simeq
\int^1_0 dx\ \varphi(x)\ (x^q-1/2), \eqno(7)
$$
where  $x=j/k.$ This indicates indeed a limit where $d(q,k)$
becomes $d(q),$ independent of $k.$ Accordingly,
$$
{dM_q \over dt} \simeq d(q) M_q. \eqno (8)
$$
Such a decoupling of eigenmodes is also found in the QCD theory
of energy-momentum
fragmentation$^{2)},$  where any moments with a (continuous) order
$q$ real and positive is an eigenmode,
in contrast however
with the present discrete
representation of fragmentation where only special values of $q$ can be
selected$^{1)}.$

\medskip
In this letter, we study a more general case where
$d(q,k)\simeq k^a d(q)$ and $a$ is any real number a priori.
This occurs in particular if
$w_{jk} \simeq k^{a-1}\varphi(j/k).$
In such a case the resulting evolution equation for moments reads
$$
{dM_q \over dt} = \sum^A_{k=1} k^{q-1} U_k\ d(q,k) \simeq\ d(q)\ M_{q+a}.
\eqno (9)
$$
Here $d(q)$ is still taken from the right-hand side of Eq.(7).
This property, Eq.(9), occurs in continuous systems$^{3)}.$
Our aim is to rather
investigate discrete fragmentations as well.

\medskip
Let us assume temporarily that Eq.(9) is an exact result. This leads to
now well-known$^{3)}$ solutions with scaling properties,
$$
M_q(t)=[s(t)]^{q-1}m_q, \ \ s(t)=\left( [s(0)]^{-a}+a \omega t \right)^{-1/a},
\ \ m_{q+a}=m_q\  (1-q)\omega/d(q), \eqno(10)
$$
where $s$ represents an average for the cluster sizes at time $t.$
The recursion relation between the coefficients $m_q$ relates
the various moments to one another via the function $d(q),$ see Eqs.(7,9,10),
which contains the dynamical information about the system. Finally
$\omega$ is an integration constant. Such scaling solutions for continuous
systems are valid only if $a$ is a strictly positive number.

\medskip
In our present study, the physical quantities under observation are the
$U_i$'s, governed by Eq.(3). The translation of Eq.(10) in terms of $U_i$
clearly reads
$$
U_i(t) \propto [s(t)]^{-1} f[i/s(t)], \eqno(11)
$$
where $f$ can be deduced from $d(q),$ but needs not be explicit in a numerical
verification of scaling. Indeed, if one solves Eq.(3) numerically,
it is sufficient to observe that, for different times $t,$ the various $U_i(t)$
show similar shapes when referred to the new scaling variable
$y \equiv i/s(t).$ Naturally, for each time $t,$
one needs a knowledge of $s(t),$ by numerical derivation if necessary.

\medskip
In this letter, we report numerical results for a two-parameter
class of models where
$$
w_{jk}=k^{a-1}[(j/k)^{-b}+(1-j/k)^{-b}]/2. \eqno (12)
$$
Such an ansatz for $w_{jk}$ was already investigated
with interesting results for the possible existence of eigenmoments$^{1)}.$
We first calculate Eq.(3) for many values
of $A,a,b,t,$ then, as a rule of thumb, discard both $U_1$ and $U_A.$
This is because an obvious peak in the $U$-histogram exists at $U_A$
for small $t$'s and at $U_1$ for large $t$'s. We are rather interested
in identifying phenomenologically $s(t)$ as the position of the
intermediate extremum (if
it exists and if it is unique) of $U_i(t)$ as a function of $i.$ The reason
for this procedure is the fact that, as already stated, the $U$-histogram
has a constant surface. An intermediate bump, or conversely
a dip, between the ``source peak'' $U_A$
and the ``sink peak'' $U_1$ is thus likely.

\medskip
Indeed, as shown by Fig.1a, where
$A=24,$ $a=0.5,$ $b=-1,$ and $t=1.2,\ 2.2,\ 3.2,$
there is an intermediate bump, which smoothly moves towards
lighter fragments when time increases. Simultaneously, the source peak
at $A=24$ decays, while the height of the fragment bump grows. The case
shown by Fig.1b is rather different. There is a
flat dip rather than a maximum, when
$A=24,$ $a=0.5,$ $b=1,$ and $t=0.28,\ 0.36.$
The flip from a source peak at
shorter times into a sink peak at larger times, see $t=0.44,$
is striking. Finally, for
Fig.1c, where $A=24,$ $a=-0.2,$ $b=-1,$ and $t=9,\ 10,$ (and $t=\ 11,$ for
comparison) we observe three maxima, namely a source peak,
a sink peak and an intermediate flat maximum.
It will be noticed that the
time scales which govern these evolutions
seem to depend strongly on $a$ and $b.$

\medskip
We now turn to the search for scaling properties. For those histograms
at time $t$ which
show an intermediate maximum $U_{max}(t)$ at some mass $i_{max}(t),$
see Fig.1a and 1c, we set $s(t)=i_{max}(t).$ Then the mass scale is defined
as $y=i/i_{max}(t),$ and we consider
${\rm Log}[U_i(t)/U_{max}(t)]$
as a function of $y.$ For those histograms which
show an intermediate minimum, we do the same, except that the reference
for scale and normalization is taken with respect to the minimum, naturally.

\medskip
This allows a comparison of histograms at different
times: all histograms reduce to 0 when $y=1.$ Finally we discard
from our analysis those histograms for which $U_i$ is monotonically increasing
or decreasing between $i=2$ and $i=A-1.$ A few results are shown on Figs.2.
Fig.2a exhibits a remarkable match of such ``renormalized'' histograms
for $A=24,$ $a=0.5,$ $b=-1,$ and $t=0.8,\ 1.4,\ 2.0.$
It must be stressed that the data for $t=0.8$ cover only the interval
$0.1 \losim y \losim 1.3,$ while those for $t=1.4$ cover only the interval
$0.2 \losim y \losim 3,$ and finally those for $t=2$ cover the interval
$0.3 \losim y \losim 4.5.$ This is because the
position $i_{max}$ decreases considerably
as a function of time. Hence the mutual continuation of the three
curves from one another is all the more striking. Another remarkable match
is seen on Fig.2b, where
$A=24,$ $a=0.5,$ $b=0.5,$ and $t=0.4,\ 0.5,\ 0.6,\ 0.7.$

\medskip
We have verified that this result does
not depend on $A,$ but only the parameters $a$ and $b,$ which must be
in a range $ a \grsim 0$
and $ b < 1.$ Indeed, as shown for instance by Fig.2c, with
$A=24,$ $a=-0.2,$ $b=-1,$ and $t=8.7,\ 9.0,\ 9.3,$
the overlap of the
renormalized histograms is less convincing. Indeed, it is seriously violated in
the region of small fragments $y \losim 0.2.$
Finally, the
results obtained when there is a minimum rather than a maximum
give no evidence for
universality of shapes in such cases, and strongly point to the
contrary, see Fig.2d with
$A=24,$ $a=0.5,$ $b=1,$ and $t=0.24,\ 0.28,\ 0.32.$

\medskip
For comparison and contrast with Fig.2a,
we show in Fig.3 the same attempt for mutual continuation with
$A=24,$ $a=0.5,$ $b=-1,$ but $t=1.8,\ 2.2,\ 2.6.$
Some success is obtained
for $y < 1,$ but the attempt fails for $y>1.$ It can be concluded that
a ``universal'' shape of the renormalized histograms is likely for intermediate
times, but not for longer times.

\medskip
It is now interesting to confront our numerical findings with the
solutions in the continuous limit$^{3)}$ where Eq.(9) is an exact property
of the system.

\medskip
i) {\it scaling }case: $a > 0, b < 1$

\smallskip
In the continuous limit, one obtains the scaling behavior described by
Eqs.(10,11). If we assume for $f$ a simple factorized form and perform the same
rescaling transformation as for the distributions $U_i,$ we get
$$
z(y) \equiv (1-b')/a'\left( 1-y^{a'}+ a'\log y \right),  \eqno (13)
$$
where the parameters $a', b'$ are to be compared with those $a,b$ of
the continuous limit
and $z $ is the scaling function emerging from the time-dependent
distributions
${\rm Log}[U_i(t)/U_{max}(t)].$ A comparison between $z$
and ${\rm Log}[U_i(t)/U_{max}(t)],$ see Fig.(4), shows that
the rescaled histogram
${\rm Log}[U_i(t)/U_{max}(t)]$ is well reproduced by $z,$ Eq.(13), with $b'=
b$,
but $a' \simeq 0.35 < a = 0.5.$  Indeed, in the continuous limit,
the scaling function $z(y), y\equiv i/s(t),$ see Eq.(11),
is known to behave as $z \propto (1-b)\log y$ for small $y$ and
$z \propto -y^a$ for large $y.$
It seems that the finite size effects do not affect the scaling properties
that we have thus observed in
the considered cases, albeit possibly modifying some of the scaling indices.

\medskip
ii) {\it shattering case}: $a < 0, b < 1$

\smallskip
In this case, scaling is approximately valid in the region $y>1$ and
strongly violated in the small $y$ region. As a matter of fact, from fitting
Fig.2b, we find empirically that the parametrization, Eq.(13),
gives a satisfactory agreement for $a' \simeq 0.1$ in the region $y>1$ but
fails for $y<1.$ Note again that $a' \ne a$. In the continuous limit,
it is known that a {\it shattering} transition$^{4)}$ takes place,
whereby a part of the system is quickly transformed into a powder of
infinitesimal constituents. In the present
discrete case, the behavior of $U$ for small values of $y$
can be traced back to a similar formation of a sector of
lightest fragments of unit mass.

\medskip
iii) {\it non-scaling (evaporation) case}: $b > 1$

\smallskip
The absence of a scaling behavior is likely due
to the presence of a minimum instead of a maximum in the
function $z(y).$ It is to be noted that the marginal value $b=1$
corresponds to the field-theoretical case mentionned in reference$^{2)}.$
However, it is not excluded that a partial scaling could be restored
by some different empirical definition of the rescaling procedure.
Nevertheless, the fact that the values $b>1$ give a strong enhancement
to the contributions of the very small (and thus very large by
symmetry) fragments at the vertex, see Eq.(7), is a hint
towards a difference between the continuous and the discontinuous cases.

\medskip
{\it Discussion and Conclusion:}

We have discovered that, at those intermediate times where
a sufficient amount of intermediary mass fragments has been generated
(and evaporation and/or shattering are not dominant), the multiplicity
histogram has a stable shape. This shape can be parametrized by simple
functions and simple parameters and evolves in time under simple {\it scaling}
rules. This result has been observed numerically for many initial masses,
even moderate ones, and a fairly general
class of models for the fragmentation vertices. The result is also true
for several excursions out of this class,
which cannot be reported here in detail.
It is thus reasonable to conjecture
that a central limit theorem, due to the semi-group, iterative nature of
the binary fragmentation cascade, is at work when the intermediary
mass fragments build a bump in the mass weighted multiplicity histogram.

We have found that our phenomenology of {\it discrete} fragmentation
shows both similarities and differences with the description$^{3)}$ proposed
for continuous systems. For example, similarities are found when
the initial mass is very large, naturally. Differences are found for
moderate initial masses. For instance we find no obvious scaling
if the histogram has a minimum (evaporation), and conversely, we find a
residual scaling for part of the histogram in the presence of shattering.
In those cases where there is an agreement between
our description and that of the continuous
limit, we find nonetheless finite size corrections in the behavior of the
scaling curve.

Besides the conjectured existence of a suitable central limit
theorem, this work opens several lines of investigation. One one hand,
those multifragmentation data which contain sizable numbers of
intermediate mass fragments should likely be analyzed in terms of scaling.
A (nuclear, atomic) clock is not necessary for dating the
corresponding histograms, since our scheme is actually time independent.
On the other hand, one should investigate whether binary$^{5)}$ cascades make
a sufficient model, or whether fragmentation is very sensitive to many-body
effects in the medium$^{6)}.$ Comparison with the properties of
percolation$^{7)}$ is also in order.

\medskip
\centerline{{\bf References}}
\medskip

\item{[1]} B.G. Giraud and R. Peschanski, {\it Phys.Lett.}
{\bf B 315} (1993) 452.

\item{[2]} G. Altarelli and G. Parisi, {\it Nucl. Phys.} {\bf 126} (1977) 297.
 V.N. Gribov and L.N. Lipatov,
{\it Sov. Journ. Nucl. Phys.}{\bf 15} (1972) 438 and 675.
For a review and references, {\it Basics of perturbative QCD} Y.L. Dokshitzer,
V.A. Khoze, A.H. Mueller and S.I. Troyan
(J. Tran Than Van ed. Editions Fronti\` eres, France, 1991.)

\item{[3]} Z. Cheng and S. Redner,
{\it J. Phys. A: Math. Gen.} {\bf 23} (1990) 1233.

\item{[4]} E.D. Mc Grady and Robert M. Ziff
{\it Phys. Rev. Lett.} {\bf 58} (1987) 892.

\item{[5]} On a phenomenological ground in relation to nuclear
multifragmentation, the problem has been
raised by: J. Richert and P. Wagner, {\it Nucl. Phys.} {\bf A 517} (1990) 299.

\item{[6]} D.H.E. Gross et al., {\it Annalen der Physik} {\bf 1} (1992) 467.

\item{[7]} X. Campi, {\it Phys. Lett.}
{\bf B 208} (1988) 351, and contributions to the
the Proceedings of Varenna 1990 and 1992
Summer Courses of the International School of Physics {\it Enrico Fermi}.
For a general review on percolation:
D. Stauffer, {\it Introduction to Percolation Theory}
(Taylor and Francis, London and Philadelphia, Penn. 1985).

\bigskip \noindent
{\bf Figure Captions}

\medskip \noindent
{\bf Fig.1a}
\smallskip \noindent
Mass weighted multiplicity $U_i$ as a function of fragment mass $i$ for
initial mass $A=24,$ and vertex parameters $a=0.5$ and $b=-1.$  See Eqs.(3,12).
Notice how the ``source peak'' $U_{24}$ decreases when time $t$ increases from
$t=1.2$ to $t=3.2.$ Simultaneously, a population of intermediate fragments
moves into lighter ones.

\medskip \noindent
{\bf Fig.1b}
\smallskip \noindent
Same as Fig.1a, but $b=1.$ In contrast with Fig.1a,
the source peak feeds directly the population of lightest fragments and even
disappears for $t=0.44.$
The process is reminiscent of evaporation.

\medskip \noindent
{\bf Fig.1c}
\smallskip \noindent
Same as Fig.1a, but $a=-0.2.$ Both source and sink peaks are present, together
with an intermediate, temporary maximum, see $t=9,10.$ A shattering phenomenon
is likely in this case.

\medskip \noindent
{\bf Fig.2a}
\smallskip \noindent
Scaling comparison of mass weighted multiplicities at various
intermediate times. For
each time we plot ${\rm Log}(U_i/U_{max})$ as a function of $y=i/i_{max}.$
The parameters are $A=24,\ a=0.5,\ b=-1,\ t=0.8,1.4,2.$
Notice how a unique scaling
curve emerges from overlapping segments of curves.

\medskip \noindent
{\bf Fig.2b}
\smallskip \noindent
Same as Fig.2a, but $b=0.5,\ t=0.4,0.5,0.6,0.7.$ Notice how intermediate
and lower mass fragments follow scaling. The population of the highest masses
deviates from the scaling curve, but its decay nonetheless contributes to
it.

\medskip \noindent
{\bf Fig.2c}
\smallskip \noindent
Same as Fig.2a, but $a=-0.2.$ Scaling is likely preserved for $y \grsim 0.8,$
and strongly violated for $y \losim 0.3.$ This indicates shattering.

\medskip \noindent
{\bf Fig.2d}
\smallskip \noindent
Same as Fig.2a, but $b=1.$ Absence of scaling when referred
to the minimum of $U_i.$

\medskip \noindent
{\bf Fig.3}
\smallskip \noindent
Same as Fig.2a, but larger time intervals. Evidence of scaling violation when
$ y > 1. $

\medskip \noindent
{\bf Fig.4}
\smallskip \noindent
Same as Fig.2a, but comparison with the theoretical (dashed line) estimate
$z(y),$ see Eq.(13).

\bye